\documentclass[12pt]{article}
\begin{document}
\date{\mbox{ }}
\title{
{\flushright{\normalsize
DESY 04-199 \\
October 2004}\\}
\vskip .3in
\textbf{Split Supersymmetry in an Orbifold GUT}\\
[5mm]}
\author{Utpal Sarkar\\
\\
{\normalsize \it
Deutsches Elektronen-Synchrotron DESY, Hamburg, Germany}\\
{{ and} \normalsize \it 
Physical Research Laboratory, Ahmedabad 380 009, India}
}
\maketitle

\thispagestyle{empty}

\begin{abstract}
\noindent

We propose a supersymmetric $SU(5)$ orbifold GUT in 5 dimensions
in which supersymmetry breaking at the boundaries give rise to 
split supersymmetry spectrum.
The extra dimension is compactified on
a $S^1/Z_2 \times Z_2^\prime$ orbifold having two fixed
points. The standard model fermions are localized
in the standard model brane, where supersymmetry
is broken by the nonvanishing of the F-term of a superfield X
at some intermediate scale. All scalar superpartners acquires
mass at this stage. The standard model gauginos and higgsinos 
remain light, since their masses are protected by R-symmetry
and induced only by loop contributions and contact interaction
in our brane. Usual features of orbifold GUTs are preserved
in this scenario and proton decay and neutrino mass problems
are solved.

\end{abstract}
\newpage
\baselineskip 18pt

{\sl Introduction:} 
The simplest grand unified theory, which accomodates the standard
model, is based on the gauge group $SU(5)$. Inspite of its simplicity,
there are several problems in the simplest version. Many extensions
were thus considered with varied consequences. To solve the gauge
hierarchy problem, supersymmetry was introduced. But the supersymmetric
models introduce more problems. Recently it has been suggested that 
if supersymmetry is broken at a very high scale, then many problems
associated with low energy supersymmetry will disappear \cite{split1,split2}. 
These split supersymmetric models will have the standard model 
fermions at low energy but their superpartners will be heavy. 
Only the superpartners of the scalar and vector particles remain
light, which can allow gauge coupling unification and solve
dark matter problem. Although there are attempts to solve 
the hierarchy problem in split supersymmetric models \cite{haba},
in most models of split supersymmetry \cite{split1,split2,split3} 
it was argued that
since we have seen fine tuning in nature like the cosmological 
constant, we should not insist on solutions of fine tuning. 

We study a
specific scheme of supersymmetry breaking in an orbifold GUT
which can give us the required spectrum of the superfields,
and in addition, maintain all the nice features of orbifold
GUTs \cite{orbi5a,orbi5b}. 
To explain the smallness of the neutrino masses we 
propose a distant breaking mechanism in orbifold GUTs in 
addition to the usual possibilities. 
The present supersymmetry breaking mechanism is similar to 
the original proposal of split
supersymmetry using Scherk-Schwarz mechanism \cite{ss} in extra 
dimensions \cite{split1}, while the distant breaking mechanism 
for neutrino mass is similar
to that applied in models of extra dimensions \cite{dist1,dist2}. 

We start with a supersymmetric SU(5) orbifold GUT in 5 dimensions
with two fixed points at ${\cal O}$ and ${\cal O}^\prime$
\cite{orbi5a,orbi5b}. In one of the branes (${\cal O}$) 
$SU(5)$ is unbroken, while 
at the other brane (${\cal O}^\prime$) only the standard 
model gauge symmetry $SU(3)_c \times SU(2)_L \times U(1)_Y$
is present.
We confine all the standard model fermions
at the standard model brane ${\cal O}^\prime$ \cite{orbi5b}. 
Supersymmetry is broken at some intermeiate energy scale in the
standard model brane (${\cal O}^\prime$) when the F-term of a scalar 
field $X$ acquires a nonvanishing value \cite{split1}. This makes all the
scalar superfields as heavy as the supersymmetry breaking
scale. However, the R-symmetry prevents the gauginos and
higgsinos from acquiring any mass. Effective higher dimensional
mass terms are allowed for the gauginos and higgsinos, which may
be induced by gravitational effects or by anomaly \cite{split1},
which are of the order of electroweak symmetry breaking scale. 
Since the fermions are confined in the standard model brane
at ${\cal O}^\prime$, fast proton decay are naturally prevented. 
The neutrino mass problem could be solved by introducing
singlet right-handed neutrinos or triplet Higgs scalar in
the standard model brane. We also propose a new
possibility of distant breaking in orbifold GUTs to explain
the smallness of the neutrino masses \cite{dist2}.

{\sl The model:}
Orbifold GUTs have been studied extensively for both
SU(5) \cite{orbi5a,orbi5b} and SO(10) \cite{orbi10} grand
unified groups. We consider a supersymmetric SU(5) 
orbifold GUT in 5 dimensions with N=1 supersymmetry.
The 5D space-time is factorized to 4D Minkowski space $M_4$ 
(with coordinates $x_\mu$, $\mu = 0,1,2,3$) and the 5th dimension
compactified on an orbifold $S^1/(Z_2 \times Z_2^\prime)$. The
circle $S^1$ with radius $R$ (with $R^{-1} \sim M_{GUT} = M_U$) will be
mod out by a discrete $Z_2$ transformation with the equivalence 
relation ${\cal P}:~y \to -y$, where $y=x_5$. We then divide 
$S^1/Z_2$ by the second $Z_2^\prime$ which acts as ${\cal P^\prime}:~
y^\prime \to - y^\prime$ with $y^\prime = y + \pi R/2$. There are
then two fixed points at the points $y=0$ and $y=\pi R /2 \equiv \ell$,
where there will be two 4-dimensional 3 branes ${\cal O}$ 
and ${\cal O}^\prime$, respectively. The action of the
two $Z_2$ and  $Z_2^\prime$ parities ${\cal P}$ and ${\cal P^\prime}$
on any generic field $\Phi(x_\mu,y)$ in the bulk will be defined
by
\begin{eqnarray}
{\cal P}&:&  \Phi(x_\mu,y) \to \Phi(x_\mu,-y) = P_\Phi ~ \Phi(x_\mu,y) 
\nonumber \\
{\cal P^\prime}&:&  \Phi(x_\mu,y^\prime) \to \Phi(x_\mu,-y^\prime) 
= P^\prime_\Phi ~ \Phi(x_\mu,y^\prime) .
\end{eqnarray}
The action of ${\cal P}$ and ${\cal P^\prime}$ give eigenvalues $\pm 1$.
The fields $\Phi_{\pm \pm} (x_\mu,y) $ with eigenvalues $\{ {\cal P},
{\cal P^\prime} \} \equiv \{ \pm,\pm \} $ will then have the following
mode expansion
\begin{eqnarray}
\Phi_{++} (x_\mu,y) &=& \sqrt{2 \over \pi R} \sum_{n=0}^\infty 
\Phi_{++}^{(2n)} (x_\mu) \cos {2 n y \over R} , \nonumber \\
\Phi_{+-} (x_\mu,y) &=& \sqrt{2 \over \pi R} \sum_{n=0}^\infty 
\Phi_{+-}^{(2n+1)} (x_\mu) \cos {(2 n +1) y \over R} , \nonumber \\
\Phi_{-+} (x_\mu,y) &=& \sqrt{2 \over \pi R} \sum_{n=0}^\infty 
\Phi_{-+}^{(2n+1)} (x_\mu) \cos {(2 n+1) y \over R} , \nonumber \\
\Phi_{--} (x_\mu,y) &=& \sqrt{2 \over \pi R} \sum_{n=0}^\infty 
\Phi_{--}^{(2n+2)} (x_\mu) \cos {(2 n+2) y \over R} ,
\end{eqnarray}
and hence only the 4D Kaluza-Klein field with eigenvalues $++$
can have massless zero mode. The fields $\Phi_{++}$ and $\Phi_{+-}$  
can be non-vanishing at the brane ${\cal O}$ at $y=0$, while
the fields $\Phi_{++}$ and $\Phi_{-+}$  
can be non-vanishing at the brane ${\cal O}^\prime$ at $y=\ell$.

In 5D the local Lorentz group is O(5). The Weyl projection
operator $\gamma_5$ is part of O(5) and hence both the left-chiral
and right-chiral fields of 4D belong to the same representation of
any 5D field. 
The $N=1$ supersymmetry in 5D will thus contain 8 real supercharges,
which in 4D will imply an $N=2$ supersymmetry.
For any realistic orbifold grand unified theory, the parity
assignment corresponding to the discrete symmetries ${\cal P}$ 
and ${\cal P^\prime}$ should reduce $N=2$ supersymmetry to 
$N=1$ supersymmetry in 4D and also break the $SU(5)$ symmetry to the
standard model gauge group $SU(3)_c \times SU(2)_L \times U(1)_Y$.
This can be achieved by the parity assignments,
\begin{equation}
{\cal P} = {\rm diag} ~\{ +1, +1, +1, +1, +1 \}
~~~ {\rm and} ~~~ {\cal P^\prime} = {\rm diag} ~\{ -1, -1, -1, +1, +1 \}.
\end{equation}
where these matrix representations of ${\cal P}$ and 
${\cal P^\prime}$ acts on the fundamental representation of
$SU(5)$. Thus all $SU(5)$ components of any multiplet will
have the same parity under ${\cal P}$, while the 
components of the $SU(5)$ multiplets that are invariant under
the standard model $SU(3)_c \times SU(2)_L \times U(1)_Y$
(denoted by the index $a$) will have opposite parity compared
to the fields belonging to the coset space
$SU(5)/SU(3)_c \times SU(2)_L \times U(1)_Y$ (denoted by the
index $\hat a$) under ${\cal P^\prime}$.

The vector multiplet of $N=2$ supersymmetry contains a vector
supermultiplet $V_a$ and a scalar supermultiplet $\Sigma_a$
of $N=1$ supersymmetry. The parity assignments are inputs in
orbifold GUTs, which determine the matter contents. One 
convenient choice for the parity operator ${\cal P}$ is even
for the vector multiplets and odd for the scalar multiplets.
The $({\cal P}, {\cal P^\prime})$ assignments for the vector
and scalar multiplets are then given by,
$$ V^a \equiv (+,+); ~~~  V^{\hat a} \equiv (+,-); ~~~ 
 \Sigma^{\hat a} \equiv (-,+); ~~~  \Sigma^a \equiv (-,-).  $$
Thus, only $V^a$ will have zero modes in the bulk. Both the fields
$V^a$ and $V^{\hat a}$ will exist in the brane ${\cal O}$ at
$y=0$ and hence all fields will experience complete $SU(5)$ 
invariance. In the brane ${\cal O}^\prime$ at $y=\ell$ the fields
$V^a$ and $\Sigma^{\hat a}$ will be present and all other fields
will experience only the standard model gauge symmetry in this brane. 
Thus $SU(5)$ symmetry will be broken to the standard model
in both the bulk and also the brane at ${\cal O}^\prime$.

In the present scenario we shall assume that the standard model
particles are localized at the ${\cal O}^\prime$ brane at $y=\ell$,
where $SU(5)$ is broken to the standard model. This is highly 
convenient for several reasons as pointed out in \cite{orbi5b}. 
The quarks and the lepton 
multiplets are just the one required by the standard model:
$q_L \equiv \pmatrix{u_L \cr d_L}$, $u^c_L$, $d^c_L$, $\ell_L
\equiv \pmatrix{\nu_L \cr e_L}$ and $e^c_L$, which remain
massless in this brane. Supersymmetry is now broken 
at a very high scale and the superpartners squarks and sleptons 
become heavy, although the higgsinos and the gauginos remain
light. 

We shall now discuss the supersymmetry breaking mechanism
in this model, which can lead to the split supersymmetry
particle spectrum. Supersymmetry is broken in the 
SU(5) invariant brane at ${\cal O}$ by the Scherk-Schwarz
mechanism \cite{ss} or by the F-component of a 
radion chiral superfield $T = r + \theta^2$ \cite{kap}. 
We follow the same procedure as that of \cite{split1,lo}.
This supersymmetry breaking background makes the minimum 
of the potential negative. The vacuum energy can then be
made to vanish by fine tuning the supersymmetry breaking
F-component of a chiral superfield $X$ in the standard model
brane ${\cal O}^\prime$. This field couples directly to the
standard model particles and hence makes all the scalar
superpartners as heavy as the supersymmetry breaking scale.
However, the gauginos and the higgsinos receive only 
anomaly mediated masses, which are of the order of 
electroweak symmetry breaking scale. 

We add a constant superpotential $$W = c M_5^3,$$ 
localized in the SU(5) invariant brane at ${\cal O}$
and break supersymmetry by the F-component of a 
radion chiral superfield $T = r + \theta^2$ \cite{kap},
which appears in the K\"ahler potential 
$$ K = M_5^3 (T + T^\dagger). $$
The tree-level effective Lagrangian for $T$
and the conformal compensator $\phi = 1 + \theta^2 F_\phi$ 
is then given by
\begin{equation}
L = \int d^4 \phi^\dagger \phi \theta K 
+ \int d^2 \theta \phi^3 W + h.c.
\end{equation}
The corresponding scalar potential
\begin{equation}
V = M_5^3 \left( r |F_\phi|^2 + F_T^* F_\phi + 3 c F_\phi + h.c
\right) 
\end{equation}
can be minimized to get the supersymmetry breaking condition
$$ F_\phi =0 ~~~~ {\rm and} ~~~~ F_T = -3c ,$$ 
with vanishing tree-level
potential. The gravitino eats up the fermionic partner of T
and becomes massive with mass $ m_{3/2} = 1/r$ (assuming $c=1$).

Although $F_\phi$ vanishes at the tree-level, at the one-loop
level it becomes nonvanishing 
$$F_\phi \sim {1 \over 16 \pi^2} { 1 \over M_5^3 r^4} \ll m_{3/2} $$
with a negative potential at the minima. To cancel the
vacuum energy, we add a chiral superfield $X$ at the
standard model brane ${\cal O}^\prime$, where all the standard
model fermions are localized. We assign a charge 2 to the field $X$
under the R-symmetry and write down the superpotential at 
${\cal O}^\prime$, given by
\begin{equation}
W = m^2 X = {1 \over 4 \pi r^2} X 
\end{equation}
and a K\"ahler potential, given by
\begin{equation}
K = X^\dagger X - {(X^\dagger X)^2 \over M_5^2} + \cdot \cdot \cdot .
\end{equation}
The minimum of the potential then breaks supersymmetry with 
$F_\phi$ nonvanishing
\begin{equation}
|F_\phi|^2 = m^4 = {1 \over 16 \pi^2 r^4}
\end{equation}
and a vanishing cosmological constant. 

The K\"ahler potential and
the superpotential results in a positive mass squared term for the
scalar component of $X$ to be $m_X^2 \sim m^4/M_5^2$; mass for the
fermionic component of $X$ to be $m_{\psi_X} \sim m^4/M_5^3$ and a $vev$ for
$X$ to be $\langle X \rangle \sim m^2 /M_5$. The effective operators
$$ \int d^4 \theta {1 \over M_5^2 } X^\dagger X Q^\dagger Q $$
would then make all the scalar partners of the fermion superfields
to be as heavy as the supersymmetry breaking scale,
\begin{equation}
m_S \sim {|F_X| \over M_5} \sim {M_5 \over M_{Pl}^4}.
\end{equation}
On the other hand, the leading effective operators contributing to
the gaugino masses,
$$ \int d^2 \theta {m^2 X \over M_5^3} W W ~~~~ {\rm and}
~~~~ \int d^4 \theta {X^\dagger X \over M_5^3} W W, $$
implies gaugino masses to be of the order of
$$M_i \sim {|F_X|^2 \over M_5^3} .$$
The origin of such effective operators could be from 
the contact interactions on the standard model brane
or could be induced by anomaly or gravitationally.
For the Higgs scalars, we assign a vanishing R-charge to
$H_u$ and $H_d$ to prevent terms like $M_5 H_u H_d$. This
makes the leading order couplings suppressed like the
effective gaugino mass term leading to $\mu B \sim |F_X|^2/M_5^3$
and $\mu \sim M_i$. 

The fundamental scale in the orbifold GUT model is the 
grand unification scale, which is of the
order of $M_5 \sim M_G \sim 3 \times 10^{13}$ GeV.
We assume the usual flat space relationship $r m_5^3 = M_{Pl}^2$. This
gives a gravitino mass of $m_{3/2} \sim 10^{13}$ GeV
and the supersymmetry breaking scale of $m_S \sim 10^9$ GeV.
All the scalar partners of the fermion superfields acquire
masses of the order of $m_S$. The gauginos and the higgsinos 
remain as light as 100 GeV. This spectrum will then be able
to allow gauge coupling unification and also a LSP dark matter
candidate. 

Since the fermions are now confined to the standard model
brane, there are no $SU(3)_c$ triplet Higgs scalars at low energy.
Fast proton decay problem is thus automatically solved in this
scenario. The quark
and lepton masses come from the usual Yukawa couplings 
\begin{equation}
{\cal L}_Y = h_u \bar q_L u_R H_1 + h_d \bar q_L d_R H_2
+ h_e \bar l_L e_R H_2
\end{equation}
in the standard model brane, where only the standard model
fermions are present and only the standard model interactions
are allowed. The smallness of the neutrino masses could have
several origin in this scenario, which we shall now discuss.

The simplest possibility would be to introduce three right-handed
neutrinos $N_i, i=1,2,3$ with large Majorana masses so
that the left-handed neutrinos receive usual see-saw mass
\cite{seesaw}. The couplings of the right-handed neutrinos
are 
\begin{equation}
L_{ss} = M_i N_i N_i + h_{\alpha i} \ell_\alpha N_i \phi + h.c.,
\end{equation}
where $\phi$ is the usual Higgs doublet of the standard model.
The Majorana mass of the physical left-handed neutrinos is then
given by
$$ m_\nu = h^\dagger M^{-1} h \langle \phi
\rangle^2 .$$
For the left-handed neutrino mass $m_\nu$ to be
in the observed range, the scale of Majorana mass of the
right-handed neutrinos should be lower than the grand
unification scale. The most natural scale for the right-handed
neutrino mass in this scenario is the supersymmetry
breaking scale of about $10^9$ GeV. 

Another possibility of explaining the small neutrino mass
is to introduce an $SU(2)_L$ triplet Higgs scalar $\xi
\equiv [1,2,-2]$ under the standard model gauge group
$SU(3)_c \times SU(2)_L \times U(1)_Y$ \cite{trip}. Right-handed 
neutrinos are not required to be present in this scenario.
If all the couplings of $\xi$ are allowed, then lepton number
is explicitly broken by its interactions
\begin{equation}
L_{trip} = f_{\alpha \beta} \ell_\alpha \ell_\beta \xi
+ \mu \xi \phi \phi + M \xi \xi
\end{equation}
where the mass of the triplet Higgs $M$ and the trilinear coupling
$\mu$ are of the same order of magnitude, which is the lepton
number violating scale. The neutrino mass matrix then
becomes
$$ m_{\nu \alpha \beta} = 
f_{\alpha \beta} ~\mu {\langle \phi \rangle^2 \over M^2} .
$$
In the present scenario this scale
could be either the supersymmetry breaking scale of $10^9$ GeV
or even the gravitino mass scale of $10^{13}$ GeV, to explain
the observed neutrino masses.

In this orbifold GUT it is
also possible to implement the distant breaking mechanism of
neutrino masses in higher dimensions \cite{dist2}.
This new possibility is similar
to that implemented in models with large extra dimensions
\cite{extra}. 
In this distant breaking mechanism for neutrino masses, we
introduce a triplet Higgs scalar $\xi$ in the standard model brane.
Lepton number is broken in the $SU(5)$ invariant brane
by the vacuum expectation value of a singlet field. Another
bulk singlet field then couples to this fields and the 
``shined'' value of the bulk singlet in our brane introduces
a small lepton number violation. 
The ``shined'' value of the bulk singlet $\eta$ in our brane
($\langle \eta \rangle$) is very small. The coupling of
this bulk singlet in our brane 
$$ L_{\eta} = \kappa \int_{{\cal O}^\prime} d^4 x \xi(x) \phi (x) 
\phi(x) \eta (x,y=\ell) 
$$
will then introduce a very tiny lepton number violating 
trilinear interaction. The neutrino mass is then given by
$$ m_{\nu \alpha \beta} = f_{\alpha \beta} \kappa \langle 
\eta \rangle {\langle \phi \rangle^2 \over M^2} .$$
Since the ``shined'' value of the bulk singlet in our brane
could be as small as of the order of eV, it is possible to
have the neutrino masses as observed, even with the triplet
Higgs scalars as light as few hundred GeV. This Higgs scalar
may then be detected in the next generation accelerator
experiments through its same sign dilepton signals. If such
signals are seen, then from a measurement of the couplings 
of the triplet Higgs scalar with the leptons it will be 
possible to determine the elements of the neutrino mass matrix,
except for an overall scale. 

{\sl Summary:}
We proposed a simple supersymmetric 
5-dimensional $SU(5)$ orbifold GUT with the 5th dimension
compactified on $S^1/Z_2 \times Z_2$. There are two fixed
points, in one of which SU(5) is broken to the standard model,
while in the other SU(5) remains invariant. The standard
model brane contains the usual fermions. Supsersymmetry
is broken by the F-component of a radion chiral superfield
in the SU(5) invariant brane, while a chiral superfield
breaks R-symmetry and supersymmetry in the standard model
brane. The scalar partners of the usual fermions then
becomes as heavy as the supersymmtry breaking scale, 
although the gauginos and the higgsinos remain as light
as the electroweak symmetry breaking scale. This can 
then allow gauge coupling unification and the lightest LSP 
could become the dark matter candidate. The SU(3) triplet
scalar becomes automatically heavy and there is no fast proton
decay problem. We also suggested few possibilities of making
neutrinos superlight, including a new distant breaking
mechanism. 

{\sl Acknowledgement:}
I would like to thank Prof. W. Buchmuller for his
invitation to DESY and acknowledge the hospitality at 
DESY, where this work was completed. Some comments of 
Prof. A. Hebecker were very helpful in improving this manuscript.

\end{document}